\documentclass[sigconf,nonacm]{acmart}
\AtBeginDocument{%
  }

\usepackage{multirow}
\usepackage{pdfpages}

\copyrightyear{2025} 
\acmYear{2025} 
\setcopyright{cc}
\setcctype{by}
\acmConference[BCB '25]{Proceedings of the 16th ACM International Conference on Bioinformatics, Computational Biology, and Health Informatics}{October 11--15, 2025}{Philadelphia, PA, USA}
\acmBooktitle{Proceedings of the 16th ACM International Conference on Bioinformatics, Computational Biology, and Health Informatics (BCB '25), October 11--15, 2025, Philadelphia, PA, USA}
\acmDOI{10.1145/3765612.3767194}
\acmISBN{979-8-4007-2200-4/2025/10}

\begin{document}

\title{Rational Multi-Modal Transformers for TCR-pMHC Prediction}

\author{Jiarui Li}
\orcid{0009-0001-1055-4424}
\email{jli78@tulane.edu}
\affiliation{%
  \institution{Department of Computer Science \\ Tulane University}
  \city{New Orleans}
  \state{Louisiana}
  \country{USA}
}

\author{Zixiang Yin}
\orcid{0009-0004-8725-1933}
\email{zyin@tulane.edu}
\affiliation{%
  \institution{Department of Computer Science \\ Tulane University}
  \city{New Orleans}
  \state{Louisiana}
  \country{USA}
}

\author{Zhengming Ding}
\orcid{0000-0002-6994-5278}
\email{zding1@tulane.edu}
\affiliation{%
  \institution{Department of Computer Science \\ Tulane University}
  \city{New Orleans}
  \state{Louisiana}
  \country{USA}
}

\author{Samuel J. Landry}
\orcid{0000-0002-4082-0543}
\email{landry@tulane.edu}
\affiliation{%
  \institution{Department of Biochemistry and Molecular Biology \\ Tulane University School of Medicine}
  \city{New Orleans}
  \state{Louisiana}
  \country{USA}
}

\author{Ramgopal R. Mettu}
\orcid{0000-0001-9479-9156}
\email{rmettu@tulane.edu}
\affiliation{%
  \institution{Department of Computer Science \\ Tulane University}
  \city{New Orleans}
  \state{Louisiana}
  \country{USA}
}

\renewcommand{\shortauthors}{Li et al.}

\begin{abstract}
T cell receptor (TCR) recognition of peptide–MHC (pMHC) complexes is fundamental to adaptive immunity and central to the development of T cell-based immunotherapies. While transformer-based models have shown promise in predicting TCR–pMHC interactions, most lack a systematic and explainable approach to architecture design. We present an approach that uses a new post-hoc explainability method to inform the construction of a novel encoder–decoder transformer model. By identifying the most informative combinations of TCR and epitope sequence inputs, we optimize cross-attention strategies, incorporate auxiliary training objectives, and introduce a novel early-stopping criterion based on explanation quality. Our framework achieves state-of-the-art predictive performance while simultaneously improving explainability, robustness, and generalization. This work establishes a principled, explanation-driven strategy for modeling TCR–pMHC binding and offers mechanistic insights into sequence-level binding behavior through the lens of deep learning.
\end{abstract}

\received{07 July 2025}
\received[revised]{10 September 2025}
\received[accepted]{27 August 2025}
\maketitle

\section{Introduction}
\begin{figure}[t]
    \centering
    \includegraphics[width=0.95\linewidth]{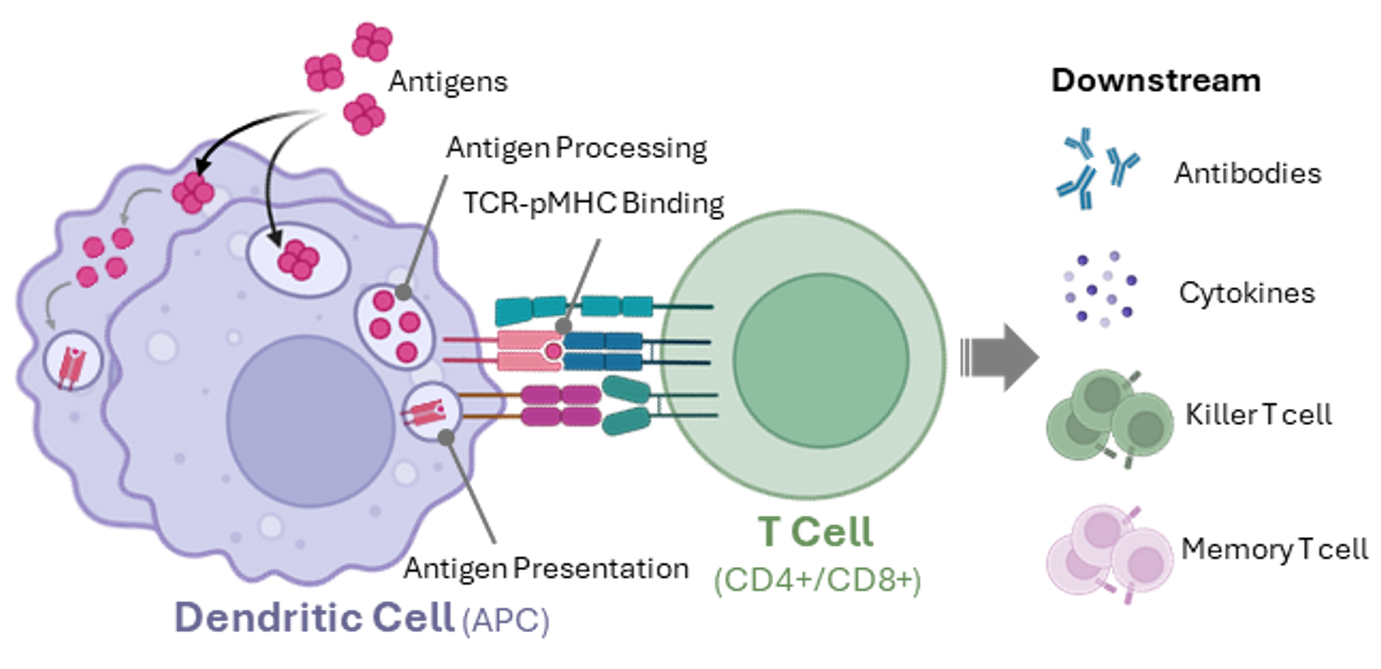}
    \caption{Binding between the peptide-MHC complex and T cell receptors is fundamental to understanding adaptive immune response and especially for developing immunotherapies (figure created in \url{https://BioRender.com}).}
    \Description{T cell receptor and peptide-MHC binding related immune activation pathways.}
    \label{fig:tcrpathway}
\end{figure}

T cells are essential components of the adaptive immune system, responsible for recognizing and responding to antigenic proteins from pathogens, such as viruses, bacteria, and cancer cells, as well as self-antigens in autoimmune contexts~\cite{joglekar2021t}. A key event in the T cell immune response is the binding between the T cell receptor (TCR) and the peptide–Major Histocompatibility Complex (pMHC), where the MHC molecule presents an antigenic peptide (i.e., epitope) on the surface of antigen presenting cells (APC). This highly specific interaction is foundational to T cell-mediated immunity (see Figure~\ref{fig:tcrpathway}) and remains a focal point in both basic immunological research and immunotherapy development. In recent years, understanding and leveraging T cell responses has become a crucial aspect of designing durable vaccines and advancing personalized cancer immunotherapies~\cite{rojas2023personalized,poorebrahim2021tcr}.

Accurate T cell response prediction requires modeling both peptide presentation and TCR recognition~\cite{peters2020t,nielsen2020immunoinformatics}. Early computational efforts emphasized peptide-MHCII binding prediction using allele-specific machine learning models~\cite{nielsen2020immunoinformatics}, such as SMM~\cite{peters2005generating,kim2009derivation}, NetMHC~\cite{lundegaard2008netmhc,nielsen2003reliable}, NetMHCpan~\cite{hoof2009netmhcpan,nielsen2007netmhcpan}, and NetMHCcons~\cite{karosiene2012netmhccons}. More recent approaches incorporate antigen processing through the Antigen Processing Likelihood (APL) algorithm~\cite{mettu2016cd4+,bhattacharya2023parallel,jiarui2024gpu,jiarui2024gpumcmc,charles2022cd4+}, which models the influence of antigen structure and its influences on peptide availability for MHCII binding.

The TCR-pMHC binding prediction problem can be formulated as a binary classification task: given a TCR sequence (in whole or selected components) and an antigenic peptide (with known MHC allele) as input, we must predict whether the TCR will bind to the pMHC complex.
Both unsupervised and supervised approaches have been explored~\cite{hudson2023can,hudson2024comparison}to analyze sequencing data from TCR-pMHC assays. Earlier unsupervised methods cluster TCR repertoires via dimensionality reduction and CDR-based similarity metrics (e.g., TCRdist3~\cite{mayer2021tcr}), without requiring binding or epitope labels (e.g., GIANA~\cite{zhang2021giana}, ClusTCR~\cite{valkiers2021clustcr}, GLIPH2~\cite{huang2020analyzing}, iSMART~\cite{zhang2020investigation}). Clusters obtained from these analyses are then used for downstream analysis~\cite{hudson2024comparison}. In contrast, more recent supervised methods leverage labeled TCR-pMHC data from resources such as VDJdb~\cite{bagaev2020vdjdb}, McPAS-TCR~\cite{tickotsky2017mcpas}, and IEDB~\cite{vita2019immune} to directly predict binding. Models such as TITAN~\cite{weber2021titan}, STAPLER~\cite{kwee2023stapler}, ERGO2~\cite{springer2021contribution}, MixTCRpred~\cite{croce2024deep}, NetTCR2.2~\cite{jensen2023nettcr}, and TULIP~\cite{meynard2024tulip} utilize deep learning architectures to achieve robust predictive performance and generalization.

\begin{figure*}[ht]
    \centering
    \includegraphics[width=\linewidth]{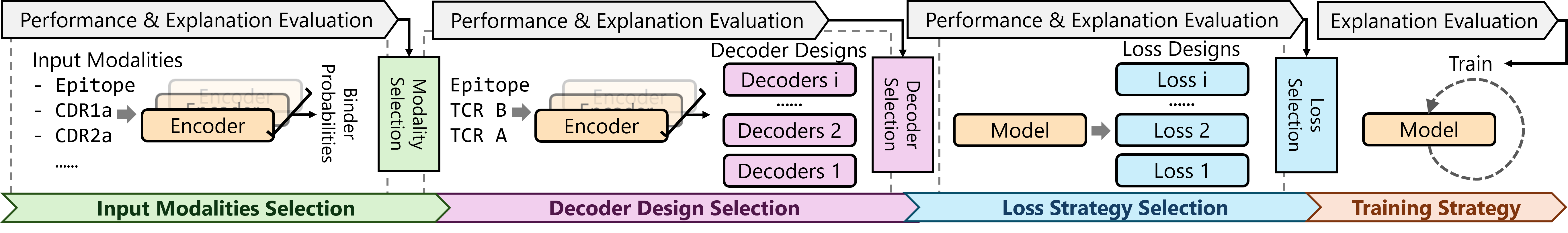}
    \caption{The development of a multi-modal transformer model for TCR–pMHC binding prediction can be systematically decomposed into four key components: input modality selection, decoder architecture design, loss function strategy, and training methodology. We employ post-hoc explainability analysis to evaluate each design choice and formulate an explanation-guided strategy across these components, ultimately resulting in a state-of-the-art predictive model.}
    \Description{The explanation guided multi-modal transformer design procedure for TCR-pMHC binding prediction.}
    \label{fig:intro:pipe}
\end{figure*}

Experimental data from TCR-pMHC binding assays may include multiple input modalities, such as full TCR sequences, complementa-rity-determining regions (CDRs), and epitope sequences. Prior studies have established CDR3 as the most critical determinant of binding~\cite{henderson2024limits}, motivating state-of-the-art models (e.g., MixTCRpred~\cite{croce2024deep}, BERTrand~\cite{myronov2023bertrand}, Cross-TCR-Interpreter~\cite{koyama2023attention}, and TULIP~\cite{meynard2024tulip}) to rely solely on CDR3 and epitope inputs. However, non-CDR3 regions have also been shown to contribute to binding prediction~\cite{henderson2024limits}. Moreover, existing models either concatenate all sequences into a single input~\cite{croce2024deep,myronov2023bertrand} or apply cross-attention exhaustively across all input pairs~\cite{meynard2024tulip,koyama2023attention}, without an explicit structural organization. 

In this paper, we present a principled approach to designing transformers with improved performance and stronger generalization for TCR-pMHC prediction by using a new method for explainability~\cite{li2025quantifying} that helps us understand the functional roles of each input and the internal dynamics of transformer-based architectures. Model explanation provides insight into why a deep learning model performs well or poorly, enabling principled analysis of how different architectural choices (i.e., cross-attention) affect model behavior. This forms the basis for a model optimization strategy driven by explainability as shown in Figure~\ref{fig:intro:pipe}. We decompose the construction of a transformer-based TCR-pMHC model into four key stages: (1) input modality selection, (2) cross-attention design for multi-modal fusion, (3) loss function strategy design, and (4) training strategy design. 

We train and test the models obtained from our approach with several TCR-pMHC datasets~\cite{bagaev2020vdjdb,tickotsky2017mcpas,vita2019immune,10x2019new,andreatta2022cd4+}. We further evaluate model generalization and explainability on the IMMREP23~\cite{morten2023immrep23} sequence benchmark and our TCR-XAI~\cite{li2025quantifying} structural benchmark. While CDR regions play the primary role in TCR-pMHC binding prediction~\cite{henderson2024limits}, our analysis demonstrates that non-CDR regions empower model encoding of the relationships between CDR regions, resulting in enhanced performance. We also explore cross-attention between CDR3b and epitope features and identify patterns of cross-attention that quantitatively improves model understanding of these modalities. Next, we demonstrate the potential that incorporating auxiliary losses and the explanation-based model training strategy can further improve generalization. We use these findings to develop a model ("EGM-2") that achieves state-of-the-art performance and generalization.  We achieve approximately 4-6\% improvements in AUC over methods such as TULIP, BERTrand and MixTCRPred in 5-fold cross validation and on the IMMREP23~\cite{morten2023immrep23} test set. Over our structure-based TCR-XAI~\cite{li2025quantifying} benchmark, we use a performance metric called binding region hit rate (BRHR) that relates model explainability with ground truth. EGM-2 achieves about a 10\% improvement in BRHR over existing methods on the TCR-XAI benchmark.

\section{Background}
In this section, we define the TCR–pMHC binding prediction problem and enumerate the relevant input modalities and define the binding prediction task. We then introduce transformer-based architectures and highlight their application in TCR–pMHC binding prediction. Finally, we describe the selected post-hoc explainable AI (XAI) techniques used to interpret these models.

\subsection{TCR–pMHC Prediction}
The TCR–pMHC binding prediction problem can be formulated as a binary classification task: given a TCR composed of alpha ($\alpha$) and beta ($\beta$) chains, an epitope $e$, and an MHC molecule $m$, the model predicts whether the pair binds (binder) or does not bind (non-binder).
The TCR chains and the epitope are proteins or peptides, typically represented as amino-acid sequences. Each TCR chain can also be described in terms of its complementarity-determining regions (CDR1, CDR2, and CDR3), as well as its variable (V) and joining (J) gene segments. The CDR regions are represented as amino acid sequences, while the V and J regions are categorical variables corresponding to specific alleles.
The classification task can be formalized as the prediction of a conditional probability:
\[
p_{\text{bind}} = P(\text{binding} \mid \alpha, \beta, e, m).
\]
If $p_{\text{bind}} > t$, where $t \in [0, 1]$ is a decision threshold, the sample is classified as a binder. Otherwise, it is classified as a non-binder.

\subsection{TCR-pMHC Prediction with Transformers}
A standard transformer architecture consists of two main components: the encoder and the decoder~\cite{vaswani2017attention}. The encoder extracts and transforms features from the inputs, while the decoder fuses these features, particularly through cross-attention mechanisms, enabling the modeling of interactions between different modalities.

TCR–pMHC binding prediction inherently involves the interaction between TCR and the pMHC complex. Consequently, encoder–decoder architectures such as TULIP~\cite{meynard2024tulip} and Cross-TCR-Interpreter~\cite{koyama2023attention} have demonstrated strong performance by explicitly modeling such interactions. However, these models typically limit their input to the CDR3 regions and epitope sequence, allowing for straightforward application of cross-attention between each input pair.
In contrast, models such as MixTCRpred~\cite{croce2024deep}, which incorporate all CDR regions along with the epitope, face increased complexity in applying cross-attention exhaustively between every pair of inputs.
While this approach yields good performance, interpretability is difficult to pinpoint the key architectural contributions.

\subsection{Post-hoc Explainability for Transformers}
As with most deep learning methods, transformers are ``black boxes'' and pose significant challenges when relating the predictions to input features. Thus post-hoc explainable AI (XAI) is an intense area of study. Initial methods focused on CNNs and other architectures~\cite{zhou2016learning,selvaraju2017grad,chattopadhay2018grad}. Recent work has developed methods for transformers~\cite{abnar2020quantifying,wu2024token,li2025quantifying,binder2016layer,achtibatattnlrp,wiegreffe2019attention,qiang2022attcat,chefer2021generic,voita2019analyzing}. TEPCAM~\cite{chen2024tepcam}, raw attention~\cite{wu2024tcr} and a method we have recently developed named QCAI~\cite{li2025quantifying}, have been used for TCR-pMHC models.  

In the context of transformer-based architectures, AttnLRP has demonstrated state-of-the-art performance for encoder-only models~\cite{achtibatattnlrp}, while QCAI has proven effective for multi-modal encoder–decoder models~\cite{li2025quantifying}. Therefore, in this work, we use these methods to interpret and analyze transformer models applied to TCR–pMHC binding prediction.
\section{Our Approach and Results}
Since this paper focuses on rational development of models, we interleave method development with experimental results, providing intermediate analyses to justify various aspects of model design. 
We decompose the transformer model design for TCR–pMHC binding prediction into four key components: (1) input modality selection, (2) cross-attention design, (3) loss function design, and (4) model training strategy design. For each component, we analyze how various design choices influence model behavior and identify the most effective configurations or improve the model, supported by explainability analyses. Based on our analyses, we obtain pan-allele TCR–pMHC binding models that achieve improved performance, explainability and generalizability over current approaches such as TULIP, BERTrand and MixTCRPred.

\subsection{Model and Training Configuration}
To control for confounding variables, we constructed all models using standard, non-pretrained BERT modules from the Huggingface transformers library. For the input modality selection, we used encoder-only transformers. For each input modality combination, an independent and identical encoder with masked language modeling loss (MLM) is applied to each input modality, and the output features are concatenated together and transformed by linear layers to predict binder or non-binder. For cross-attention design, the input features are processed by encoders following the same configuration as in input modality selection. For  loss strategy design, each auxiliary loss is linked to the independent linear layers transforming the transformers' output features.

Each encoder and decoder module consists of two hidden layers with 128-dimensional hidden states and a single attention head to minimize computational overhead. All models were trained for 500 epochs using the AdamW optimizer with a learning rate of $1E^{-4}$. When considering the IMMREP and TCR-XAI test sets, we examine an explanation-based training strategy (Section~\ref{result:ts}). All training was performed on a machine equipped with two NVIDIA A2000 GPUs and two Intel E5 CPUs.

\subsubsection{Datasets}
To train and evaluate model performance, we collected a positive dataset following the procedure described in MixTCRpred~\cite{croce2024deep}, aggregating TCR-pMHC binding data from both \textit{Homo sapiens} and \textit{Mus musculus} across multiple sources: VDJdb~\cite{bagaev2020vdjdb}, McPAS-TCR~\cite{tickotsky2017mcpas}, IEDB~\cite{vita2019immune}, 10X Genomics~\cite{10x2019new}, Andreatta et al.~\cite{andreatta2022cd4+}, and Zander et al.~\cite{zander2022delineating}. Negative samples were generated by pairing TCRs with non-binding pMHCs, maintaining a 1:1 ratio of negative to positive examples. For each epitope, we sampled an equal number of negative pairs to ensure class balance. Model performance was evaluated using 5-fold cross-validation.

\subsection{Evaluation Metrics}
For each model, we first evaluate its performance using 5-fold cross validation on the compiled training dataset.
Then, to assess generalization, we train models on the full training set and evaluate them on IMMREP23~\cite{morten2023immrep23}, a public benchmark for TCR–epitope specificity prediction that includes peptides unseen during training. To probe the internal mechanisms of the models and understand how they interpret input features, we applied post-hoc explainability methods: AttnLRP~\cite{achtibatattnlrp} for encoder-only models and QCAI~\cite{li2025quantifying} for encoder–decoder models. We only use binder classification loss to generate explanations for encoder-only models and use the training loss for encoder-decoder models. To generate attention weights, we consider all "not available" values (NA) as 0, and apply a smoothing operation using a convolution operation with core $[1/3, 1/3, 1/3]$ to tolerate one residue offset. 

\subsubsection{TCR-XAI Benchmark}
Explanation quality was assessed using our recently developed TCR-XAI benchmark~\cite{li2025quantifying}, which quantifies how well model-generated importance scores align with structural ground truth.
It consists of 274 high-resolution crystal structures of TCR-pMHC complexes from the STCRDab~\cite{leem2018stcrdab} and TCR3d 2.0~\cite{lin2025tcr3d} datasets. The availability of these structures gives us a means to objectively evaluate both accuracy and explainability.
We use the Binding Region Hit Rate (BRHR)~\cite{li2025quantifying} to assess explanation quality. This score reflects how effectively the explanation method identifies actual binding residues based on structural proximity. Intuitively, BRHR compares top-ranked residues by explanation score against top interacting residues by distance.
To calculate BRHR, we choose a percentile threshold $t \in (0, 1]$ and select the top $t$ fraction of residues with the highest importance scores $\mathbf{S}$. A residue is counted as a \textit{hit} if its structural interaction distance also falls within the top $t$ fraction. For each sequence type of each sample that is predicted as a binder by a given model, we compute the individual hit rate, then average these values across the dataset (TCR-XAI) to produce the final BRHR for that model. 
In this paper, we use $t=0.25$ to decide whether a residue is correctly identified as involved in binding; this is the most strict threshold that produces at least one binding region in every sample in the TCR-XAI set. We have tested other thresholds exhaustively and find similar results for all experiments in this paper.

\subsection{Input Modality Selection}
Most existing models use only the CDR3 regions and epitope sequences as input~\cite{meynard2024tulip,koyama2023attention}. Although the CDR3b region is widely acknowledged as a key determinant of TCR–pMHC interaction, recent studies suggest that non-CDR3 and even non-CDR regions may also contribute meaningfully to TCR–pMHC binding~\cite{henderson2024limits}. Therefore, the contributions of other TCR components have not been thoroughly investigated. 

To address this gap, we conduct two sets of experiments: one to assess how different CDR regions affect model behavior, and another to examine the impact of both CDR and non-CDR regions on TCR–pMHC binding prediction performance.

\subsubsection{CDR Regions}
\begin{table}[ht]
    \caption{The ROC-AUCs of transformer models evaluated across different combinations of input modalities.
    Boldfaced values (0.902 and 0.755) denote the best 5-fold and test performance; the need to include both CDR3b and epitope sequences is evident.}
    \label{tab:results:rep:cdrs_tcrs_rocauc}
    \centering
    \begin{tabular}{lll}
        \toprule
         Input Modalities &  5-Fold & Test \\
         \midrule
         CDR3b only & 0.488$\pm$0.007 & 0.505\\
         Epitope only & 0.507$\pm$0.005 & 0.500\\
         CDR3b + Epitope & \textbf{0.704$\pm$0.008} & 0.517\\
         All CDR3s + Epitope & 0.756$\pm$0.010 & 0.607\\
         All CDRs + Epitope & 0.847$\pm$0.004 & 0.694\\
         \midrule
         TCR A + All CDRbs + Epitope& 0.899$\pm$0.005 & 0.751\\
         TCR B + All CDRas + Epitope & 0.893$\pm$0.004 & \textbf{0.755}\\
         TCRs + Epitope & 0.867$\pm$0.003 & 0.753 \\
         TCRs + All CDRs + Epitope & \textbf{0.902$\pm$0.003} & 0.738\\
         \bottomrule
    \end{tabular}
\end{table}
To evaluate how various CDR regions influence model behavior, we trained five encoder-only models using different combinations of CDRs and epitope. This design mirrors input modality combinations commonly adopted in prior work. The input modalities and the related representative TCR-pMHC prediction models are as follows:
CDR3b only (e.g., TCRdist3~\cite{mayer2021tcr}, GIANA~\cite{zhang2021giana}),
Epitope only,
CDR3b + Epitope (e.g., BERTrand~\cite{myronov2023bertrand}, epiTCR~\cite{pham2023epitcr}),
CDR3s + Epitope (e.g., TULIP~\cite{meynard2024tulip}, TSpred-Attention~\cite{kim2024tspred}),
and All CDRs + Epitope (e.g., MixTCRPred~\cite{croce2024deep}, NetTCR-2.2~\cite{jensen2023nettcr}, TSpred-CNN~\cite{kim2024tspred}).
For each configuration, we used separate encoder modules to extract features from each modality. The resulting embeddings were then concatenated and passed through a linear classification layer to predict binding outcomes.

As shown in Table~\ref{tab:results:rep:cdrs_tcrs_rocauc}, the ROC-AUC results from both 5-fold cross-validation and independent test set evaluation indicate that incorporating additional input modalities than only using CDR3b and epitope improves model performance from 0.704 to 0.847 and generalization ability from 0.517 to 0.694. Notably, the model requires at least both the epitope and CDR3b as inputs to develop a valid understanding of TCR–pMHC binding. 

\begin{figure}[ht]
    \centering
    \includegraphics[width=0.95\linewidth]{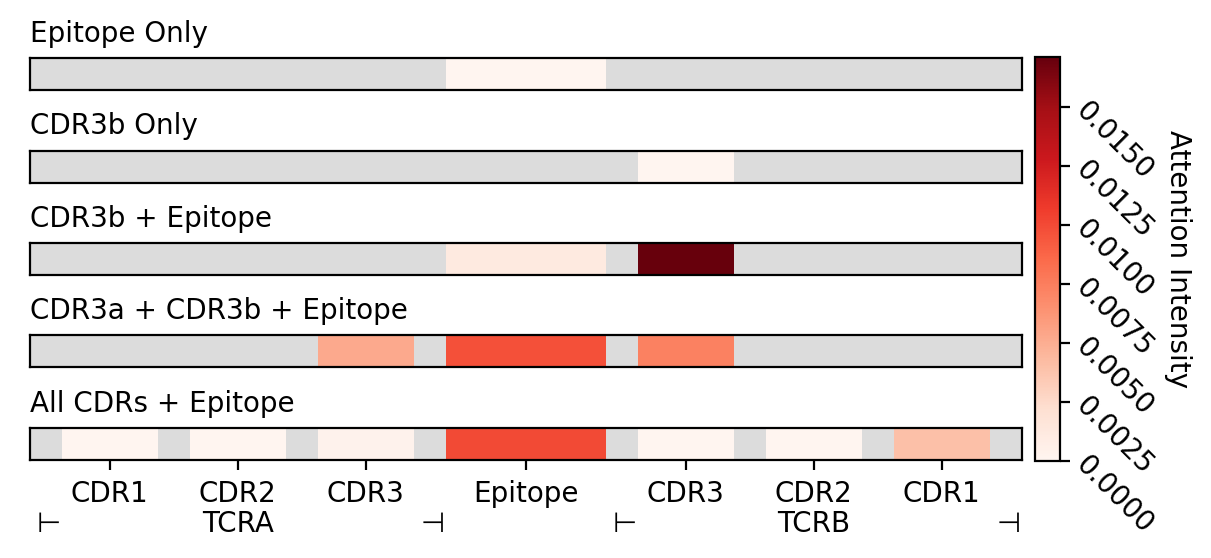}
    \caption{The sample-wise average attention intensities across different TCR and epitope regions from transformers with various CDR input modalities.}
    \Description{The sample-wise average attention intensities of different regions of TCRs and epitope.}
    \label{fig:result:cdr_attn_weight}
\end{figure}

An interesting observation is that while combining CDR3b and epitope inputs leads to a substantial performance improvement over using either modality alone in 5-fold cross-validation, the generalization ability improves by only 1.7\%. However, when using both CDR3a and CDR3b as input,  generalization ability improves more than 10\%. To understand this discrepancy, we analyzed attention intensity and model explanation results (Figure~\ref{fig:result:cdr_attn_weight}). 
The analysis reveals that the inclusion of epitope and CDR3b as inputs lead to model bias attention on CDR3b and limited model interpretability. However, using both CDR3a and CDR3b as input significantly enhances the model’s ability to interpret the CDR3b region. This suggests that while epitope and CDR3s are essential for the model to learn meaningful binding representations, the inclusion of additional CDRs (particularly CDR3a) enables the model to better learn the binding pattern of the epitope and improves mutual understanding between CDR3a and CDR3b under binding scenarios. 

However, when all CDR regions are used as inputs, the average attention intensity on CDR regions decreased and explanation average BRHR slightly decline about 0.01.
The model appears to struggle with encoding the relationship between CDR regions effectively. Although this configuration achieves improved performance and generalization, the explainability, particularly in terms of epitope attention and explanation BRHR, declined by 0.08. These findings suggest that if we can better structure the input from all CDR regions, it may be possible to further improve both model performance and generalization.

\subsubsection{Full TCR Sequences}
To improve the model’s ability to organize input from the CDR regions and to investigate the contribution of non-CDR regions to TCR–pMHC binding prediction, we extend the input modalities to include full TCR sequences. These models follow the same configuration as those used in the CDR region experiments. The input modalities for each configuration are as follows:
(1) All CDRs + Epitope, (2) TCR A sequence + All CDRb regions + Epitope, (3) TCR B sequence + All CDRa regions + Epitope, (4) Full TCR sequences + Epitope, and (5) Full TCR sequences + All CDR regions + Epitope.

As shown in Table~\ref{tab:results:rep:cdrs_tcrs_rocauc}, while the 5-fold performance shows only moderate improvement to 0.867 from 0.847, the generalization ability increases substantially to 0.753 from 0.694. Notably, using only the full TCR A or TCR B chain still achieves ROC-AUCs of 0.751 and 0.755 respectively on the independent test set. However, when both full TCR sequences and all CDR regions are included as inputs, the model achieves an ROC-AUC of 0.902 on the 5-fold validation but a lower ROC-AUC of 0.738 on the independent dataset, indicating overfitting. These findings suggest that incorporating either TCR A or TCR B full sequence is sufficient to enhance both performance and generalization. To better understand the underlying mechanism, we further analyzed the explanation quality and attention intensity of these models.

\begin{figure}[ht]
    \centering
    \includegraphics[width=0.95\linewidth]{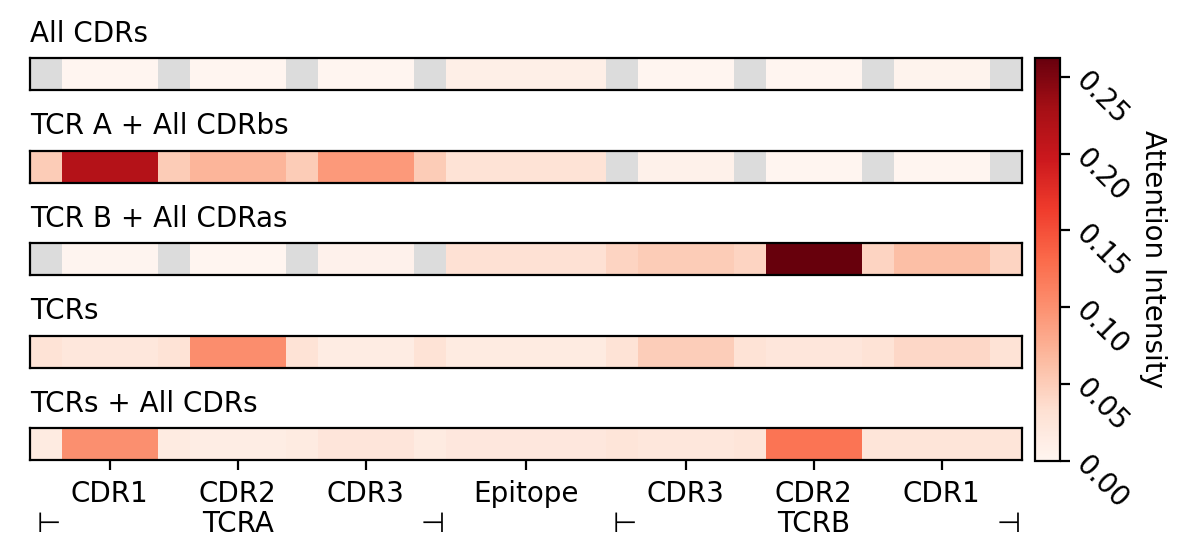}
    \caption{The sample-wise average attention intensities across different regions of TCRs and the epitope from transformers with different TCR input modalities.}
    \Description{The sample-wise average attention intensities of different regions of TCRs and epitope.}
    \label{fig:result:tcr_attn_weight}
\end{figure}

As shown in Figure~\ref{fig:result:tcr_attn_weight} and Table~\ref{tab:results:rep:cdrs_tcrs_rocauc}, incorporating full TCR chains allows the model to assign it higher attention intensity and better understand the epitope-TCR interaction with 0.04 BRHR improvement. In particular, the full TCR B chain enables the model to capture TCR B-epitope interaction more effectively with an 0.06 BRHR increase. However, the model gains a worse understanding of interactions between TCR A and TCR B with decreases of 0.02 and 0.05 in BRHR, respectively. 

Consistent with previous findings, these results indicate that although full TCR sequences help the model learn more about the epitope and the model primarily relies on CDR regions for accurate TCR–pMHC binding prediction. However, the increased sequence length and complexity make it difficult for the model to process and organize all the information effectively. Therefore, it is crucial to develop strategies that structure and prioritize input information to improve model explainability and generalization.

\subsection{Cross-Attention for Multi-Modal Fusion}
\begin{figure*}[ht]
    \centering
    \includegraphics[width=0.95\linewidth]{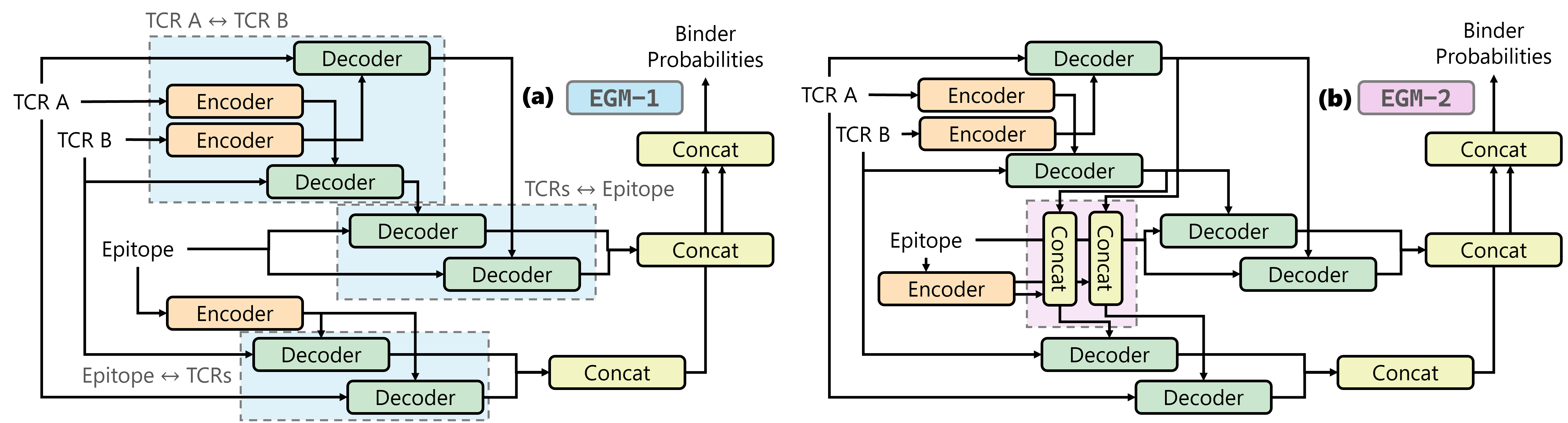}
    \caption{The architectures of the explanation-guided models (EGM). EGM-1 (a) includes additional decoders to capture internal TCR interactions first and independent decoders to capture epitope-TCR interactions. EGM-2 (b) is an improved version of EGM-1 with extra information from the other TCR chain and was developed by analyzing  explanability.}
    \Description{The designed encoder-decoder model based on previous explanation analysis with TCRs and epitope as input.}
    \label{fig:result:designed_model_v1}
\end{figure*}

To address complexity of solving feature relationships and dependencies, explicitly modeling the interactions between input features using a structured design can lead to improved performance and generalization. One effective approach to achieve this is by employing a decoder with cross-attention~\cite{vaswani2017attention}, which enables controlled and explainable information flow between different input components. Before constructing such a model, it is essential to understand the cross-attention mechanism within the decoder architecture.

\subsubsection{Analysis of Decoder Cross-Attention}

\begin{table}[ht]
    \centering
    \caption{The ROC-AUCs of transformer models with various cross-attention designs for epitope and CDR3b. The cross-attention $a \to b$ only preserves information from $b$, because ROC-AUC of $a\to b + a$ is about 0.72 while $a\to b (+ b)$ is near random.}
    \label{tab:results:crossattn_mechanism_aucroc}
    \begin{tabular}{ll}
    
        \toprule
         Cross-Attentions &  5-Fold \\
         \midrule
         Epitope$\to$CDR3b & 0.520$\pm$0.008\\
         Epitope$\to$CDR3b + CDR3b & 0.522$\pm$0.006 \\
         Epitope$\to$CDR3b + Epitope	& \textbf{0.732$\pm$0.006}\\
         CDR3b$\to$Epitope & 0.484$\pm$0.004\\
         CDR3b$\to$Epitope + CDR3b &	\textbf{0.718$\pm$0.007}\\
         CDR3b$\to$Epitope + Epitope & 0.478$\pm$0.004\\
         CDR3b$\leftrightarrow$Epitope	& \textbf{0.718$\pm$0.007}\\
         \bottomrule
    \end{tabular}
\end{table}

\begin{table}[ht]
    \centering
    \caption{The Binding Region Hit Rate (BRHR) for transformers with different cross-attention designs between the epitope and CDR3b.  These results highlight the ability of cross-attention to directionally enhance the model's understanding of inter-modality interactions.} 
    \label{tab:results:crossattn_mechanism_hrbr}
    \begin{tabular}{lllll}
    \toprule
     &  & Epitope & CDR3b & CDR3b \\
     &  Interact& \,\,\,\,\,\,\,\,$\downarrow$ & \,\,\,\,\,\,\,\,$\downarrow$ & \,\,\,\,\,\,\,\,+ \\
    Modalities & with & CDR3b & Epitope & Epitope \\
    \midrule
    \multirow[t]{4}{*}{Epitope} & TCR B & 0.7080 & \textbf{0.7638} & 0.6636 \\
     & CDR1b & 0.6722 & \textbf{0.7248} & 0.6441 \\
     & CDR2b & 0.6657 & \textbf{0.7610} & 0.6924 \\
     & CDR3b & 0.7211 & \textbf{0.7886} & 0.6805 \\
    \cline{1-5}
    CDR3b & Epitope & \textbf{0.7369} & 0.6122 & 0.7286 \\
    \bottomrule
    \end{tabular}
\end{table}

The decoder is composed of multiple layers, each containing a self-attention (encoder) layer and a cross-attention layer. The cross-attention mechanism allows one input embedding (the query) to attend to and extract information from another input or a concatenated set of inputs (the keys and values). This enables explicit modeling of interactions between distinct input modalities. However, it remains unclear whether cross-attention truly enhances the model’s understanding of interactions between the query and the attended inputs, and what specific information is ultimately propagated through this mechanism. To investigate this, we design an experiment using CDR3b and epitope sequences as inputs, aiming to elucidate how cross-attention captures and represents their interaction.

We investigate the directional behavior of the decoder’s cross-attention by applying it between two input modalities: CDR3b and epitope. Specifically, we construct models in which one input modality serves as the query to attend to the other (e.g., CDR3b $\to$ epitope), and analyze performance both when using only the decoder output, where we denote $a \to b$ as using $a$ as the query to attend to $b$. As shown in Table~\ref{tab:results:crossattn_mechanism_aucroc}, using cross-attention alone between CDR3b and epitope yields similar performance to using either input independently around 0.5, essentially random guessing, suggesting that cross-attention tends to preserve information primarily from only one input.

To determine which input's information is retained, we include the original features of either the query or the attended input in the final prediction layer. We observe that combinations like epitope $\to$ CDR3b (with CDR3b features) and CDR3b $\to$ epitope (with epitope features) significantly improve performance to 0.732 and 0.718 respectively, whereas using the query features yields lower performance. This indicates that, in the cross-attention $a \to b$, the decoder primarily retains information from $b$, the attended modality.

To further investigate the explainability of cross-attention, we analyze explanation quality using binding region hit rate (BRHR) as shown in Table~\ref{tab:results:crossattn_mechanism_hrbr}. For epitope $\to$ CDR3b, the model shows improved understanding from CDR3b to epitope with BRHR achieving 0.7369, and conversely, for CDR3b $\to$ epitope, the model better understands epitope to CDR3b interactions with BRHR reaching 0.7638. This directional explainability suggests that cross-attention $a \to b$ enhances the model’s ability to capture interactions from $b$ to $a$.
Based on these observations, we propose to leverage directional cross-attention in encoder–decoder architectures to explicitly guide and enhance interaction modeling between TCR and pMHC, thereby improving both performance and generalization capability.

\subsubsection{Explanation-Guided Cross-Attention Design} 

As demonstrated in our exploration of input modalities, incorporating full TCR chains enhances the model’s understanding of both the epitope and the TCR itself. Therefore, we adopt TCR A, TCR B, and the epitope as input modalities for the design of our encoder–decoder architecture. 
The simplest approach to constructing such a model (EGM-0) is to apply direct cross-attention from one modality to the other two, following the design principle of TULIP~\cite{meynard2024tulip}. This design enables the model to enhance its representation of a given modality by attending to complementary contextual information from the others.
\begin{table}[ht]
    \centering
    \caption{ROC-AUCs of explanation-guided models (EGM-1, EGM-2) versus baselines~\cite{myronov2023bertrand,meynard2024tulip,croce2024deep}. Our models consistently outperform baselines in 5-fold cross-validation and test set evaluation, demonstrating enhanced predictive performance and generalization.}
    \label{tab:result:design_rocauc}
    \begin{tabular}{lll}
        \toprule
         Models &  5-Fold & Test \\
         \midrule
         CDR3s + Epitope (BERTrand~\cite{myronov2023bertrand}) & 0.704$\pm$0.008 & 0.517\\
         CDR3s$\leftrightarrow$Epitope (TULIP~\cite{meynard2024tulip})	& 0.803$\pm$0.002 & 0.566 \\
         All CDRs + Epitope (MixTCRPred~\cite{croce2024deep}) & 0.847$\pm$0.004 & 0.694\\
         \midrule
         EGM-0 (TCRs$\leftrightarrow$Epitope)	& 0.879$\pm$0.006 & 0.750 \\
         \midrule
         \textbf{EGM-1} & 0.885$\pm$0.003 & 0.760\\
         \textbf{EGM-2} & \textbf{0.888}$\pm$0.002 & \textbf{0.765}\\
         \bottomrule
    \end{tabular}
\end{table}

\begin{table}[ht]
    \caption{The Binding Region Hit Rate (BRHR) for explanation-guided models. EGM-1 demonstrates improvement on inter-TCR interaction understanding and EGM-2 increases understanding among all interactions.}
    \label{tab:results:design_hrbr}
    
    \begin{tabular}{lllll}
    \toprule
     & Interact & EGM-0 & EGM-1 & EGM-2 \\
    Modalities & with & (TCR$\leftrightarrow$Epitope) & & \\
    \midrule
    \multirow[t]{2}{*}{Epitope} & TCR A & 0.7019 & 0.7456 & \textbf{0.7821} \\
     & TCR B & 0.6394 & 0.7207 & \textbf{0.7341} \\
    \cline{1-5}
    \multirow[t]{2}{*}{TCR A} & Epitope & 0.7981 & 0.7320 & 0.7404 \\
     & TCR B & 0.7309 & \textbf{0.7750} & \textbf{0.8024} \\
    \cline{1-5}
    \multirow[t]{2}{*}{TCR B} & Epitope & 0.6457 & 0.6798 & \textbf{0.8413} \\
     & TCR A & 0.6459 & \textbf{0.6809} & \textbf{0.7742} \\
    \bottomrule
    \end{tabular}
\end{table}

As shown in Table~\ref{tab:result:design_rocauc}, applying direct cross-attention between TCR sequences and epitopes enhances 5-fold cross-validation performance to 0.879 but does not improve generalization ability, which keep 0.75. To investigate this, we leverage model explanations. According to Table~\ref{tab:results:design_hrbr}, the model exhibits limited improved understanding of the interaction from TCR B to TCR A, which is smaller than 0.05. In addition, its understanding of interactions from TCR A and TCR B to epitope declines from 0.9109 to 0.7981 and from 0.9476 to 0.6457 respectively. These findings suggest two potential directions for further model improvement: (1) enhancing the model's ability to capture interactions from epitope to TCRs, and (2) improving its understanding of the interactions from TCRs to the epitope.

We find that EGM-0 exhibits insufficient understanding of the interaction between TCR A and TCR B, as indicated by BRHR scores of 0.65 (TCR B $\to$ TCR A) and 0.73 (TCR A $\to$ TCR B). Following the first strategy, we designed the initial version of our explanation-guided model (EGM-1) architecture, as illustrated in Figure~\ref{fig:result:designed_model_v1}a. To enhance the model’s understanding of epitope-TCR interactions with extended representational capacity, we first apply cross-attention between TCR A and TCR B chains. Subsequently, the epitope sequence is used to query each TCR chain independently. This enables epitope-TCR interactions to be processed with independent decoders. Also, to further expand the representational capacity and improve predictive performance, we employ separate decoders to perform cross-attention from both TCR A and TCR B to the epitope.

As shown in Table~\ref{tab:result:design_rocauc}, this architecture effectively improves both cross-validation and generalization performance to 0.885 and 0.76 respectively. Explanation based analysis further reveals that the model develops a stronger understanding of TCR A–TCR B, epitope-TCR A, and epitope-TCR B with 4\%, 3.5\%, and 8\% BRHR improvement. However, due to the use of independent decoders for TCR-to-epitope attention, the model’s ability to capture joint TCR–epitope interactions is reduced.

Although EGM-1 improved modeling of TCR inter-chain interactions, its understanding of epitope–TCR interactions remains limited: the BRHR from TCR B to the epitope is only 0.68, significantly lower than other interaction BRHR scores. Building upon EGM-1 and incorporating the second improvement strategy, we designed a second version of the model, EGM-2, illustrated in Figure~\ref{fig:result:designed_model_v1}b. In EGM-2, to enhance the model’s understanding of epitope interactions, we modify the decoder responsible for cross-attention from the TCR chains to the epitope. Specifically, we integrate additional features from the complementary TCR chain during cross-attention. This design allows the model to contextualize each TCR chain’s interaction with the epitope in the presence of the other chain’s information, thereby fostering a more comprehensive understanding of TCR–epitope binding patterns. According to the Table~\ref{tab:result:design_rocauc}, it achieves 0.765 ROC-AUC on test dataset. With respect to explainability, as shown in Table~\ref{tab:results:design_hrbr}, it demonstrates 10\% BRHR improvement for all interactions in average. In particular, the BRHR of interaction between TCR B and epitope increases 20\% to 0.8413.

\subsection{Loss Strategies}
\begin{figure*}[ht]
    \centering
    \includegraphics[width=0.9\linewidth]{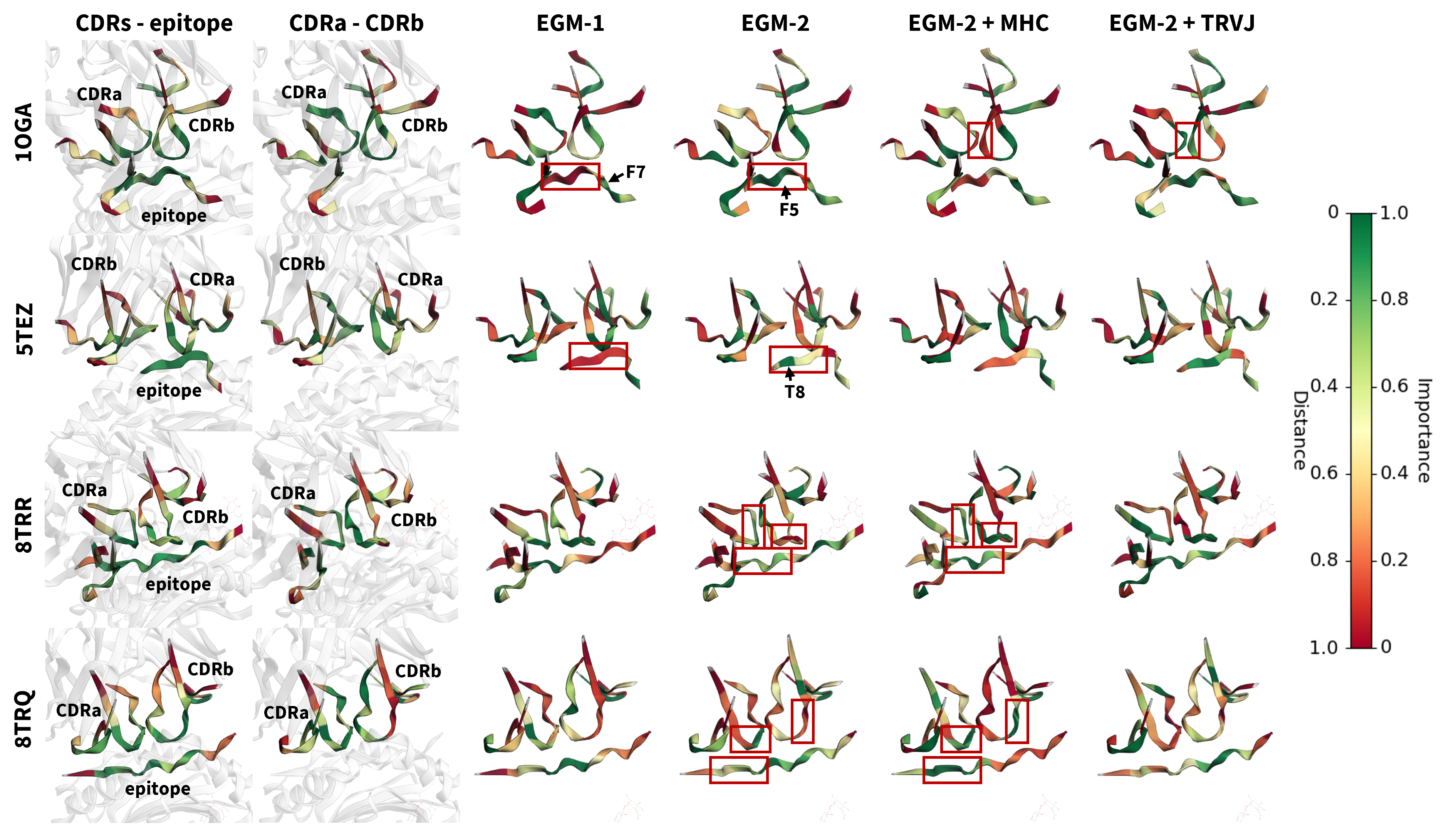}
    \caption{Two TCR-pMHC structural case studies from TCR-XAI.  \texttt{1OGA} and \texttt{5TEZ} are structures as examples for an influenza epitope and the same TCR (top two rows), while \texttt{8TRR} and \texttt{8TRQ} are examples for a rheumatoid arthritis epitope bound to two distinct TCRs (bottom two rows). For \texttt{1OGA} and \texttt{5TEZ}, compared to EGM-1, EGM-2 shifts attention from residue F7 to F5 and maintains attention on T8, while in \texttt{1OGA}, TRVJ loss improves model ability for analyzing inter-TCR relationships. For \texttt{8TRR} and \texttt{8TRQ}, MHC loss shifts attention from epitope center to CDR3a side and yields enhanced understandings of TCR-epitope interaction.}
    \Description{TCR-XAI case studies on systems for EGM-1, EGM-2, MHC loss, and TRVJ loss.}
    \label{fig:result:case_study}
\end{figure*}
Loss strategies are a significant component of transformer model design, guiding model optimization during training and influencing the representational capacity of the model. For all previously discussed models, we employed two types of loss functions: masked language modeling (MLM) loss and binder classification loss. To identify the most effective loss strategy for TCR–pMHC prediction, we conducted a two-step investigation: (1) evaluating the role of MLM loss, and (2) exploring potential auxiliary losses to further enhance model performance. We used EGM-2 to explore loss strategies.

\begin{table}[ht]
    \centering
    \caption{Binding Region Hit Rate (BRHR) of EGM-2 under different loss strategies. MLM loss improves modality-level understanding, while MHC and TRVJ allele classification losses enhance interpretability for epitope recognition and inter-TCR interactions respectively.}
    \label{tab:result:loss_brhr}
    \begin{tabular}{llllll}
    \toprule
     \multirow[t]{2}{*}{Loss} & MLM & - & \checkmark & \checkmark & \checkmark \\
     & Auxiliary & - & - & MHC & V/J  \\
    Modalities & Interact with &  & & &  \\
    \midrule
    \multirow[t]{2}{*}{Epitope} & TCR A  & 0.7533 & 0.7821 & \textbf{0.8166} & 0.7581 \\
     & TCR B & \textbf{0.8075} & 0.7341 & 0.5817 & 0.7018 \\
    \cline{1-6}
    \multirow[t]{2}{*}{TCR A} & Epitope & \textbf{0.7498} & \textbf{0.7404} & 0.6307 & 0.5508 \\
     & TCR B & 0.6841 & 0.8024 & 0.8086 & \textbf{0.8321} \\
    \cline{1-6}
    \multirow[t]{2}{*}{TCR B} & Epitope & 0.7606 & \textbf{0.8413} & \textbf{0.7910} & 0.6869 \\
     & TCR A & 0.6627 & \textbf{0.7742} & 0.7140 & \textbf{0.7515} \\
    \bottomrule
    \end{tabular}
\end{table}

\subsubsection{Masked Language Modeling Loss}
The masked language modeling (MLM) loss masks parts of the input and uses cross-entropy to evaluate how well the model can recover the masked tokens.
In our previous models,
we applied MLM loss to both encoders and decoders. To assess how MLM loss affects the binder classification task, we trained the designed model with encoders regularized by MLM loss, but decoders optimized solely with the binder classification loss.
Although removing MLM loss from the decoders little decrease 5-fold validation and generalization ROC-AUC within 0.001 and 0.05 respectively.
Explanation analysis in the Table~\ref{tab:result:loss_brhr} indicates that removing decoder MLM loss substantially impaired the model's understanding of the interaction from epitope to TCR A and from TCR B to epitope and TCR A. 
These results demonstrate that MLM loss enhances the decoder's ability to understand the input data, ultimately improving model robustness.

\subsubsection{Auxiliary Loss}
Based on the dataset composition, we identified two potential auxiliary losses: (1) MHC categories and alleles (MHC loss), and (2) V and J region alleles of TCRs (TRVJ loss). The MHC category can be formulated as a binary classification task (MHC-I vs. MHC-II), while both MHC alleles and the V/J region alleles of TCRs can be framed as auxiliary multi-class classification tasks.
Incorporating these auxiliary losses slightly improves ROC-AUC on the test dataset about 0.005, which is minor and not significantly different ($p>0.5$) to the model without auxiliary losses. However, according to the explanation evaluation in the Table~\ref{tab:result:loss_brhr}, the TRVJ loss improves model understanding between TCRs and MHC loss enhances model's understanding from epitope to TCR A and TCR B to epitope.
This suggests that these auxiliary objectives mediate models' behavior and affect the way models capture the interaction between TCRs and epitope.
These experiments demonstrate that the MLM loss improves model understanding among all input modalities, auxiliary classification loss for MHC enhances the model's understanding of TCR A and epitope interaction, and V/J alleles auxiliary classification loss boosts the model's explanation between TCRs.

\subsubsection{Case Studies}
To demonstrate the explainability difference between loss strategies, we conducted two groups of case studies shown in Figure~\ref{fig:result:case_study}. The first case study considers two TCR-pMHC complexes from the MHC-I pathway for an influenza epitope; structures \texttt{1OGA}~\cite{stewart2003structural} and \texttt{5TEZ}~\cite{yang2017structural} capture two binding registers for the same TCR. 
In \texttt{1OGA}, EGM-2 shifts attention from residue F7 to F5, which is closer to CDR loops. In \texttt{5TEZ}, EGM-2 maintains attention on T8 near the CDR3a loop. However, for EGM-2 with TRVJ loss, we note that the model highlights the contacts between CDR3 regions correctly only in \texttt{1OGA}. This is because our analysis is restricted to CDR regions, which represent only a subset of the full TCR sequence.

The second case study considers two distinct TCRs (\texttt{8TRR} and \texttt{8TRQ}~\cite{loh2024molecular}) for an epitope from a self-antigen associated with rheumatoid arthritis in the MHC-II pathway.
For these two structures, EGM-2 with MHC loss shifts its attention from the epitope center (near by CDR3a and CDR3b) to the CDR3a side, enhancing understanding of epitope–TCR A interaction while reducing insight into epitope–TCR B interaction. In addition, it focuses attention from the flank of CDR3 loops to the center of CDR3 loops, which are the key regions contacting the epitope. 

\subsection{Training Strategy}
\label{result:ts}

For all previously discussed models, we trained for 500 epochs and selected the model with the lowest training loss as the best model. However, this approach risks overfitting to the training data. Determining an effective training stopping criterion remains a critical challenge in neural network optimization. A common practice is to stop training based on minimal training loss or peak performance on a validation set. However, the former may result in overfitting, while the latter can suffer from poor generalization ability if the validation set fails to represent the full data distribution. We find that explanation-based metrics could help to address this.

\begin{figure}[ht]
    \centering
    \includegraphics[width=0.9\linewidth]{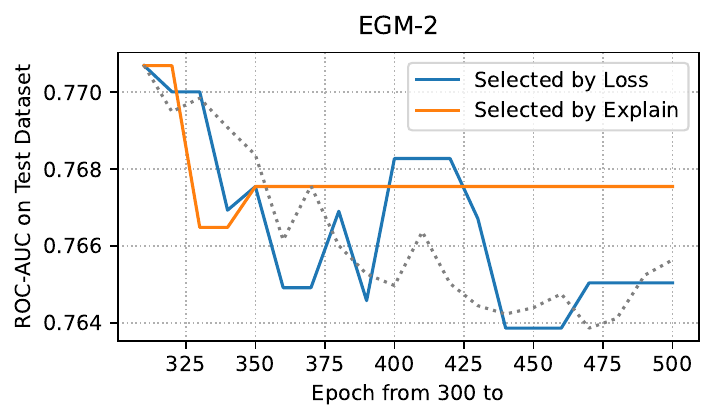}
    \caption{ROC-AUC on the independent test set for models selected by minimal loss or explanation quality (epochs 300–500). From epoch 350, explanation-based selection yields more stable and better generalizing models. }
    \Description{Comparison of ROC-AUC for models selected using loss-based and explanation-based strategies on the test set, evaluated from epoch 300 to 500.}
    \label{fig:result:train_selected}
\end{figure}

We began evaluating model selection strategies from epoch 300, by which time the model was close to convergence. Two strategies were compared: (1) Loss-based: where the model with the lowest training loss was selected, and (2) Explanation-based: where the model demonstrating the best explanation quality was chosen. Explanation quality was assessed by evaluating the model's understanding of four bidirectional interactions: from TCRs to epitope and from epitope to TCRs. As shown in Figure~\ref{fig:result:train_selected}, prior to epoch 350, both strategies yielded models with comparable ROC-AUC performance on the test set. However, after epoch 350, when the loss is stable, the explanation-based strategy consistently selected models with better generalization performance and robustness.

These findings suggest that models exhibiting stronger explainability in terms of biologically meaningful interactions are more likely to generalize well. Consequently, it is possible to use explanation-based metrics as indicators to stop training to obtain models with better generalization ability. 
\section{Conclusion}

In this paper, we have systematically deconstructed transformer model design for TCR–pMHC binding prediction into four critical components: (1) input modality selection, (2) model architecture, (3) loss strategies, and (4) training methodology. Through comprehensive experimentation guided by analysis of model explainability, we have identified and validated novel model designs that achieve state-of-the-art performance for TCR-pMHC prediction. 


Our findings underscore three core principles: (1) full TCR sequences enable cross-chain contextual learning; (2) directional attention mechanisms are key to explainable binding prediction; and (3) explanation-guided training strategies foster better generalization. 
These findings suggest a way to build self-explainable models for TCR-pMHC binding. We believe that the understanding gained in analyzing directional attention mechanisms can enable us to build models based on \textit{concepts}~\cite{koh2020concept} (i.e., an explainable sub-structure in the model) that effectively capture the interaction between modalities. In addition, auxiliary losses can serve as a regularization technique for concept learning.

\textbf{Code and Data Availability}: The code, models, and data introduced in this paper can be found at \url{https://github.com/Tulane-Mettu-Landry-Lab/tcr-rational}.
\begin{acks}
We thank the anonymous reviewers, area chairs, and program chairs for their valuable feedback. This work was supported by National Institutes of Health (U54-CA260581) ``Tulane University COVID Antibody and Immunity Network (TUCAIN)'', AWS Cloud Research Credits, and the Harold L. and Heather E. Jurist Center of Excellence for Artificial Intelligence at Tulane University.
\end{acks}


\bibliographystyle{ACM-Reference-Format}
\bibliography{reference}

\includepdf[pages=-]{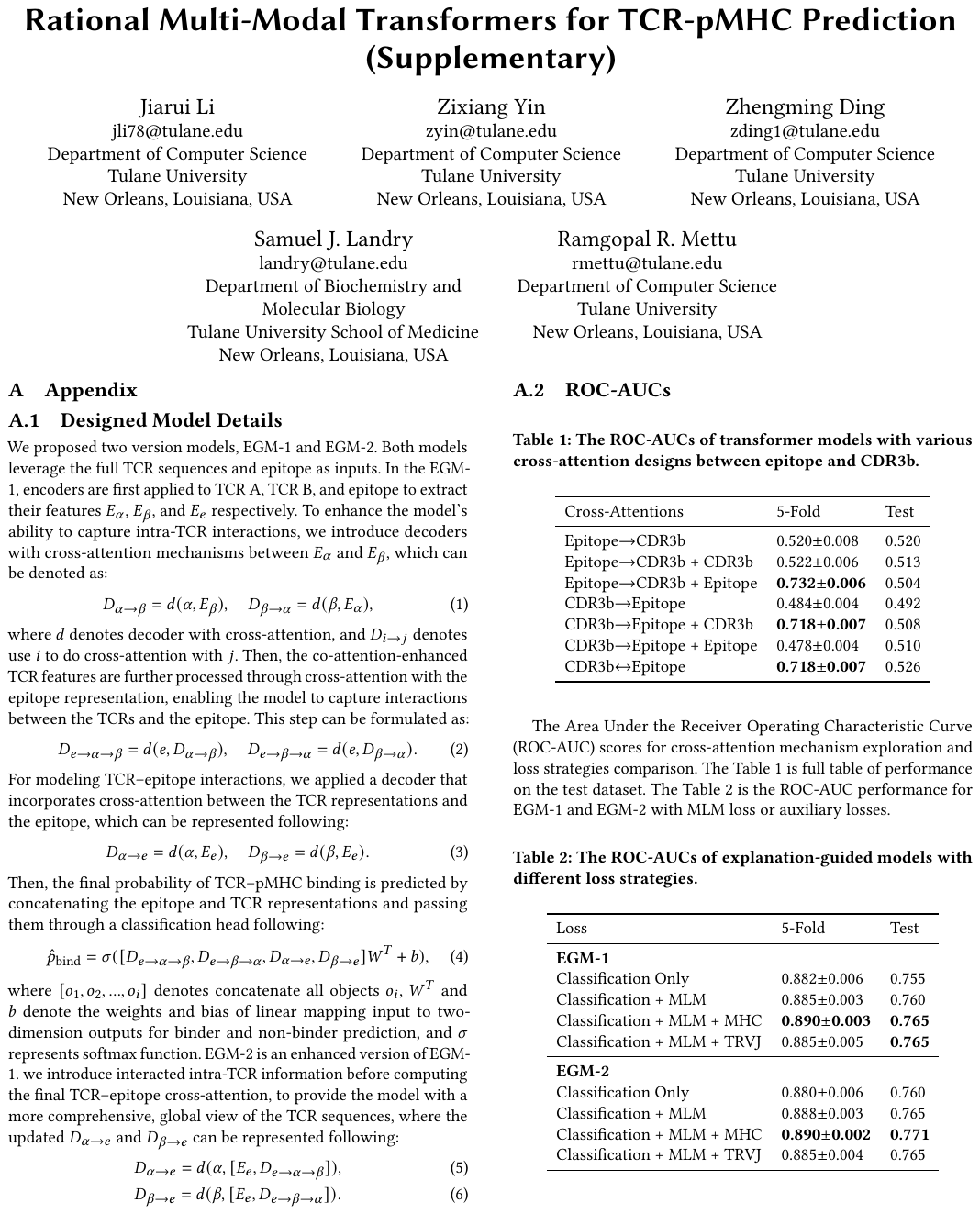}

\end{document}